\documentstyle[prl,aps,twocolumn]{revtex}

\begin{document}
\draft
\title{Observation of the $e/3$ fractionally charged Laughlin quasiparticles}
\author{L. Saminadayar and D. C. Glattli}
\address{Service de Physique de l'\'{E}tat Condens\'{e}, \\
CEA/Saclay, F-91191 Gif-sur-Yvette Cedex, France}
\author{Y. Jin and B. Etienne}
\address{Laboratoire de Microstructures et Micro\'electronique, \\
C.N.R.S., B. P. 107, F-92225 Bagneux Cedex, France}
\date{\today}
\maketitle

\begin{abstract}
The existence of fractional charges carrying the current is experimentally
demonstrated. Using a 2-D electron system in high perpendicular magnetic
field we measure the shot noise associated with tunneling in the fractional
quantum Hall regime at Landau level filling factor 1/3. The noise gives a
direct determination of the quasiparticle charge, which is found to be $%
e^{*}=e/3$ as predicted by Laughlin. The existence of $e/3$ Laughlin
quasiparticles is unambiguously confirmed by the shot noise to
Johnson-Nyquist noise cross-over found for temperature $\Theta=e^{*}V_{ds}/2k_{B}$.
\end{abstract}

\pacs{PACS numbers : 73.40.Hm, 05.30.fk, 71.27.+a, 72.70.+m}

\narrowtext
Can fractional charges carry the current in a conductor? Up to now, there
was no evidence of such phenomenon. Usual metals are known to form Fermi
liquids with quasiparticles of charge $e$. Low dimensional systems are
believed to offer a richer spectrum of excitations. Indeed, fractional
charges have been predicted for commensurate charge density waves in one
dimensional systems \cite{Schr}, and for two-dimensional electron systems
(2DES) \cite{Laug} in high perpendicular magnetic field when the fractional
quantum Hall effect occurs \cite{Tsui}. In this letter, we report
experimental evidence of charges $e/3$ carrying the current. The observation
is done in the Fractional Quantum Hall (FQH) regime at Landau level filling
factor $\nu=1/3$.

2-D electrons in high magnetic field give rise to degenerate Landau Levels
(LL) with one state per flux quantum $\phi _{0}=h/e$ in the plane. For
integer LL filling factor $\nu =n_{s}/n_{\phi }$, the cyclotron or the
enhanced Zeeman gap gives rise to the integer quantum Hall effect~\cite
{Klitzing} IQHE ($n_{s}$ and $n_{\phi }=eB/h$ are the electron and quantum
flux density) \cite{Pran}. The simplest elementary excitation is an electron
removed from the highest occupied LL, leaving a hole having the size of a
flux quantum and a unit charge $e$. At fractional filling factor $\nu =1/q$, 
$q$ odd, a gap $\Delta \simeq e^{2}/\epsilon l_{c}$ also opens resulting
from the interactions \cite{Yosh} ($l_{c}=(\hbar /eB)^{1/2}$). This is the
FQH effect. Laughlin has proposed \cite{Laug} that an elementary excitation
can be realized by introducing a flux quantum $\phi _{0}$ in the collective
wavefunction. As there is one electron for $q$ flux quanta, the so-called
Laughlin quasiparticle has fractional charge $e^{*}=e/q$. Extensions of the
Laughlin approach to higher rational fractions $\nu =p/q$ \cite{Hald}
explain many bulk transport properties but so far no direct experimental
evidence for the bulk Laughlin quasiparticles have been found. The
quasiparticles in the bulk have been mostly probed using thermal activation.
The prefactor of the activated conductivity has shown a striking~\cite{Clar}
relation to the quasiparticle charge but is not fully understood\cite{Poly}.
Comparison of the chemical potential jump at fractional $\nu $ obtained from
capacitance measurements with the activation energy may also determine $e^{*}
$ but is not straightforward \cite{Doro}. An alternative way to determine $%
e^{*}$ uses the edge rather than the bulk quasiparticle properties. At the
edge, the Landau levels bent by the confining potential cross the Fermi
level and form edge channels. The gapless modes at the Fermi energy, the
edge states, provide the dominant conduction in low disorder samples~\cite
{Edge}. For integer $\nu $ the edge channels are chiral Fermi liquids with
quasiparticles of charge $e$. For $\nu =p/q$, fractional edge channels
similarly form. They are chiral Luttinger liquids \cite{Wen,Mill} with
fractionally charged quasiparticles generalizing the bulk Laughlin
quasiparticles. Attempts to measure $e^{*}$ have used the Aharonov-Bohm
period of the conductance \cite{Simm,Gold,Fran}. In a recent beautiful
experiment using an anti-dot at $\nu =1/3$ \cite{Gold}, the period of the
polarization charge on the control back-gate was found accurately $e/3=e^{*}$%
. In a similar report, it has been argued that {\it equilibrium}
conductance measurements mostly probe the fractional filling of the
ground state \cite{Fran}. An experiment specifically sensitive to the charge
carrying the current was thus needed.

The following experiment is based on a {\it non-equilibrium} property, the
shot noise, which probes the granularity of the quasiparticles carrying the
current. According to Schottky \cite{Scho}, a poissonian uncorrelated flow
of carriers generates current fluctuations. The noise power $S_{I}$ is
directly proportional to the carrier charge. A quasiparticle tunneling
through the $\nu =1/3$ quantum Hall fluid is expected for weak coupling
between opposite 1/3 edge channels. Indeed, the tunneling density of states
of quasiparticles diverges at the Fermi energy while that of electron
vanishes \cite{Wen}. The tunneling, or backscattering, current is : $%
I_{B}=(e^{2}/3h)V_{ds}-I$. $V_{ds}$ is the voltage difference between ideal
contacts connecting the edges, $I$ the total current. At temperature $\Theta
=0$ and for weak coupling $I_{B}\ll I$, the Schottky formula gives \cite
{Kane,Fend,Cham}: 
\begin{equation}
S_{I}=2(e/3)I_{B}
\end{equation}
The noise is thus a direct measure of $e^{*}=e/3$. If instead electrons were
to tunnel the noise power would be $2eI_{B}$. When the quasiparticle
chemical potential difference $(e/3)V_{ds}$ becomes smaller than the
temperature a cross-over to Johnson-Nyquist noise is expected \cite
{Kane,Fen2} ($I_{B}\ll I$): 
\begin{equation}
S_{I}=2\left[ \left( e/3\right) I_{B}\coth \left( \frac{e^{*}V_{ds}}{%
2k_{B}\Theta }\right) -2k_{B}\Theta \frac{dI_{B}}{dV_{ds}}\right]
+4k_{B}\Theta \frac{dI}{dV_{ds}}
\end{equation}
The characteristic voltage is three times that expected for electrons. The
second part of the right hand side identifies to the Johnson-Nyquist noise
at zero bias. Note the differential conductance in the expression. For
non-Fermi liquids $I_{B}(V_{ds})$ is always non-linear \cite{Wen}.
Expressions for $dI/dV_{ds}$ and $S_{I}$ for the ideal case of a single
tunneling impurity are given in ref.~\cite{Wen,Fend,Kan2}. In real samples
with smooth edges and where tunneling is induced by the smooth potential of
a Quantum Point Contact (QPC) deviations from the calculable ideal case may
be expected but the general predicted features remain. We emphasize that the
exact edge state dynamics (the way $I_{B}$ varies with $V_{ds}$) is less
essential here as {\it in the weak backscattering limit the noise must be
given by Eq.(1)}.

In this limit, the shot noise measurements reported here agree with Eq.(1),
bringing evidence that $e/3$ charge quanta do carry the current at $\nu =1/3$%
. Our observation also confirms the Johnson-Nyquist to shot noise cross-over
given by Eq.(2) . The samples are GaAs/Ga(Al)As heterojunctions with low
density $n_{s}=0.94X10^{15}m^{-2}$ high mobility $100m^{2}V^{-1}s^{-1}$ $%
100nm$ deep 2DES. Six wide ohmic contacts with increased perimeter length
provide ideal contacts in the fractional regime. Using electron beam
lithography technique metallic gates are evaporated at the center of the
wide Hall bar mesa to define $275nm$ wide QPC. The QPC locally creates a ($%
\simeq 150\phi _{0}$) wide $\nu =1/3$ region upon applying a negative
voltage on both gates while keeping a constant filling factor $\nu _{L}=2/3$
in the mesa, see Fig.$1a$ . The $1/3$ state is signaled by a $e^{2}/3h$
conductance plateau when sweeping the gate voltage. A quasiparticle
tunneling through the $1/3$ state is induced upon applying a slightly more
negative voltage. The saddle shape QPC potential combines with the weak
random potential to give few tunneling paths whose interference leads to
conductance oscillations at the end of the $1/3$ plateau. Using the
independent control of the gates we can {\it laterally shift the constriction%
} to tune the tunnel coupling of a particular conductance peak. Each peak is
found to reach a maximum value remarkably equal to $e^{2}/3h$, see Fig. $1b$%
. This observation tells us that {\it tunneling is coherent} and that a $%
{\it 1/3}${\it \ FQHE state is still formed} for this gate voltage range.
This is an important check without which the following noise results would
have been questionable~\cite{Note}. Finally, looking at the differential
conductance we see that the overall conduction is restored at finite d.c.
bias voltage $V_{ds}$, in qualitative agreement with chiral Luttinger
models. Fig. $1c$ shows the $dI/dV_{ds}$ characteristics for a series of
gate voltage decribing the left wing of the resonance of Fig.$1b$ . The
global features, similarly observed in many samples, are consistent with a
singular density of state being a decreasing function of the energy with
respect to the Fermi surface as expected for quasiparticle tunneling \cite
{Wen}. We are thus in the good regime to detect $e/3$ charges; a detailed
analysis of the non-linear transport will be given elsewhere. For the
following noise measurements we only need to know $I_{B}$ and keep~$I_{B}\ll
I$.

Noise measurements use the correlation method decribed in \cite{Kuma,Glat}
for the observation of the Pauli suppression of fermion shot noise, but here
the sample is voltage biased, see Fig. $1a$. The voltages $V_{5,6}$ and $%
V_{3,2}$ are separately measured by two ultra low noise amplifiers and a
spectrum analyzer calculates the cross-correlation spectrum. This technique
removes from the detected signal the uncorrelated amplifier voltage noise
and the noise of the leads and contacts. The cross-correlation spectrum $%
S_{V_{5,6}V_{3,2}}$ contains the physical shot noise contribution $S_{I}$
plus some white noise sources of the circuit $S_{V_{4,1}}$, $S_{I_{3}}$, and 
$S_{I_{6}}$ : $%
S_{V_{5,6}V_{3,2}}=R_{H}^{2}S_{I}+(1-R_{H}dI/dV_{ds})^{2}(S_{V_{4,1}}+R_{H}^{2}(S_{I_{3}}+S_{I_{6}})) 
$ where $R_{H}=3h/2e^{2}$ is the quantized Hall resistance of the mesa. The
circuit noise sources require to keep $dI/dV_{ds}$ constant to reliably
extract $S_{I}$. This is done with a $0.2\%$ accuracy for each series of
noise measurements. The amplifier gains known to $0.5\%$ allow for accurate
determination of $S_{I}$. Finally Johnson-Nyquist noise measurements for
different temperatures at fixed conductance {\it in the fractional regime}
provide an {\it absolute} calibration as in \cite{Kuma}.

The result of a series of current noise power measurements versus the
backscattering current $I_{B}$ at $\Theta =25mK$ is shown in Fig. $2$ . The
noise measured in the $4-8KHz$ frequency range is white. The background
noise $\simeq 5.32$ $10^{-28}A^{2}/Hz$ is due to the circuit noise. The
error bars represent the statistical accuracy expected for $1500s$
acquisition time. $I_{B}$ can be varied by changing either the dc bias $%
V_{ds}$ or the tunnel coupling with gate voltage. In order to keep $%
dI/dV_{ds}=G_{diff}$ constant and follow the path A shown in Fig. $1b$ and $%
1c$ both the bias ($40\mu V$ to $78\mu V$) and the gate voltage ($-170.5$ to 
$-178.5mV$) are varied. The backscattering current is obtained within $5\%$
accuracy by measuring the d.c. voltage $V_{B}=V_{3,2}$ or $V_{5,6}$ across
the QPC, using $I_{B}=(2V_{B}-V_{ds})e^{2}/3h$. The ``reflexion
coefficient'' $R=I_{B}3h/e^{2}V_{ds}$ is kept small for weak backscattering.
It increases with $I_{B}$ from $4\%$ to $35\%$. The linear variation of the
noise with $I_{B}$ tells us that we do observe shot noise associated with
backscattering. We can compare the rate of noise variation with that given
by Eq.(1) (dashed line). {\it The agreement with the prediction of Laughlin
quasiparticle tunneling is excellent}. Electron tunneling would have given a
very different result (dotted line). Electron Shot-Noise is found for
similar conductance $G=0.32e^{2}/h$ at lower field in the Integer Quantum
Hall regime ($\nu _{L}=4$ in the leads), inset of figure $2$.\ The data
agree with the electron theory for a lowest Landau level transmission $0.32$ 
\cite{Les1}.

How this remarkable result is robust against parameter changes? Figure $3a$
shows the current noise versus $I_{B}$ for two different $G_{diff}$ (path B
and C). The noise also compares well with that expected for $e/3$ charges
except for the points at high bias where the backscattering is no longer
weak and less noise is found. A good agreement is also found for a different
tunneling regime obtained by detuning a resonance ( Fig. $1d$, path D : $%
V_{ds}=78\mu V$ to $175\mu V$ and gate voltage $-161$ to $-177mV$). The
result is also robust against temperature change as shown by the series E
corresponding to the tunneling conditions of Fig. $1b$ but at $\Theta =150mK$%
. Finally, a room temperature thermal cycling changes the resonance shape but
not the noise results.

How to take into account the deviations for large $R$? As long as electron
tunneling do not start to compete with quasiparticle tunneling, we may
expect a decrease of noise when $R$ increases. Indeed, the tunneling events
are no longer poissonian as the exclusion statistics and the interactions
correlate the quasiparticles. If they were fermions a noise reduction $(1-R)$
would occur \cite{Kuma,Les1,Rez}. It is not legitimate \cite{Fen2} but
nevertheless tempting to plot the noise data as a function of $I_{B}(1-R)$
(open circles of Fig. $2$ and $3$). Within experimental accuracy, the simple 
$(1-R)$ reduction factor accounts well for the data but slightly
overestimates $e^{*}$. The least square linear fit gives $e^{*}=0.38$, $0.36$%
, $0.35$, and $0.36$ for A, B, C, and D.

The final check to confirm our observation of $e/3$ Laughlin quasiparticles
is the cross-over from Johnson-Nyquist to shot noise at $e^{*}V_{ds}/2=k_{B}%
\Theta $. Fig.~$4$ shows measurements at $\Theta =134mK$ and low bias. Here,
the bias voltage $V_{ds}$ varies from $13\mu V$ to $140\mu V$ and $%
G_{diff}=0.26e^{2}/h$. The nearly linear noise variation at high bias,
consistent with Eq. ($1$), saturates at low bias. The arrow, indicating when $%
e^{*}V_{ds}=2k_{B}\Theta $, is well in the cross-over region. Comparison
with Eq.(2) (solid curves) shows a remarkable agreement (the experimental
variation of $V_{ds}(I_{B})$ is used). The dashed curve describing the
thermal cross-over for charge $e$ does not fit the data.

In conclusion, using shot noise, we have brought evidence of $e/3$ Laughlin
quasiparticles carrying the current through the $1/3$ FQH state. The result
is robust against various tunneling and temperature conditions and the
Johnson-Nyquist to shot noise cross-over confirms the nature of the
quasiparticles. At larger backscattering a noise reduction factor similar to
that expected for fermions accounts for the observation but slightly
overestimates the noise. After submission of this work, we became aware of a
similar observation by another group~\cite{Picci}.

We warmly thank A. Kumar for participating at the early stage of the
experiment and P. Jacques for his precious technical contribution. Useful
discussions with Th. Martin, H. Saleur, I. Safi, M.\ B\"{u}ttiker, C. de C.
Chamon, M. H. Devoret, and V. Pasquier are acknowledged.

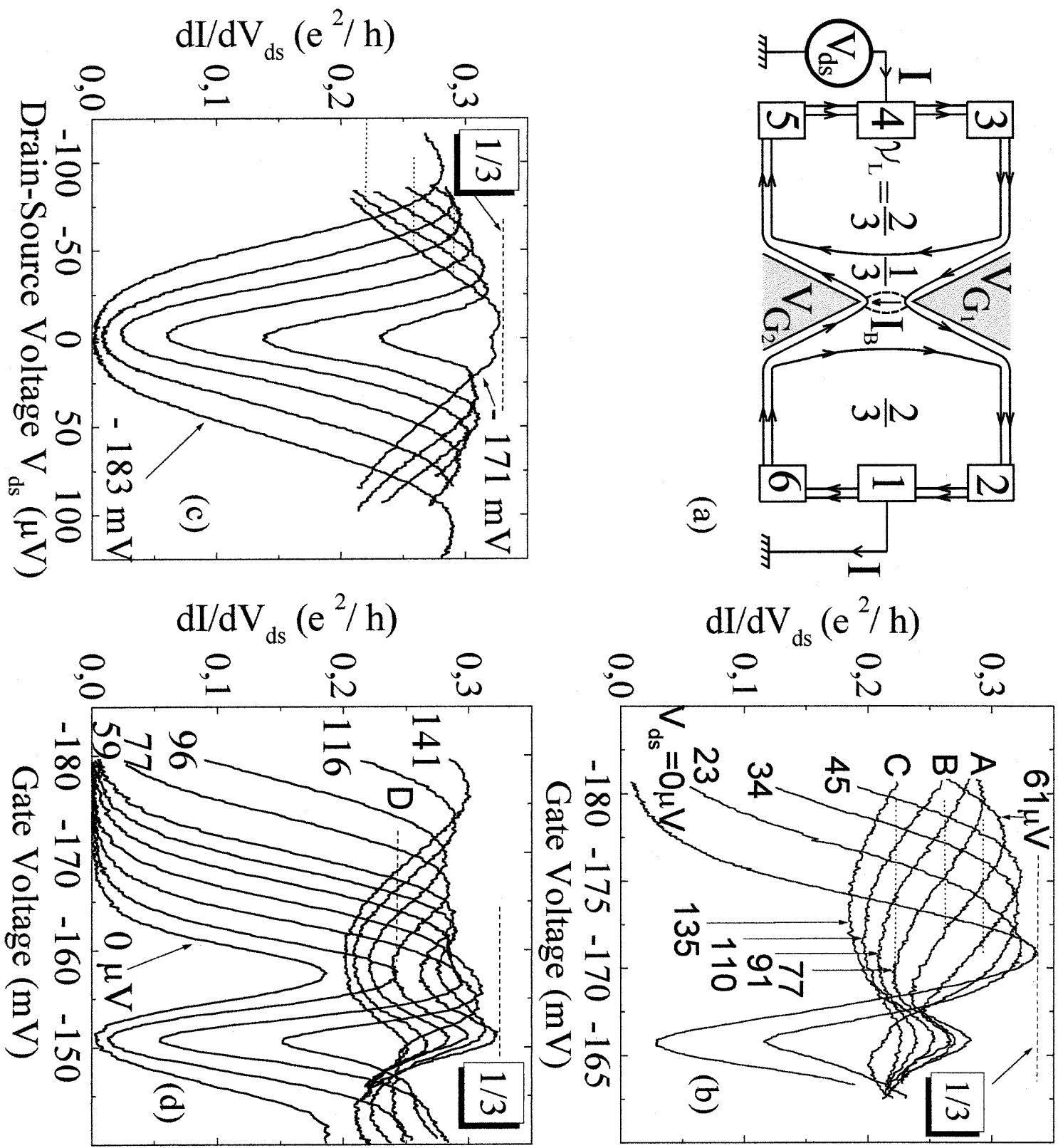
\begin{figure}[tbp]
\caption{(a) Schematic picture of the Hall sample. (b) Differential
conductance versus gate voltage for different bias $V_{ds}$ for a tuned
resonance at $\nu=1/3$. (c) $dI/dV_{ds}$ versus $V_{ds}$ for different gate
voltages ($2mV$ steps) describing the left wing of the tuned resonance. (d)
Untuned resonance:~$dI/dV_{ds}$ versus gate voltage for different $V_{ds}$ ($%
10$$\mu$$V$ steps from $0$ to $59$$\mu$$V$).~All data are taken at $25mK$.}
\label{fig1}
\end{figure}
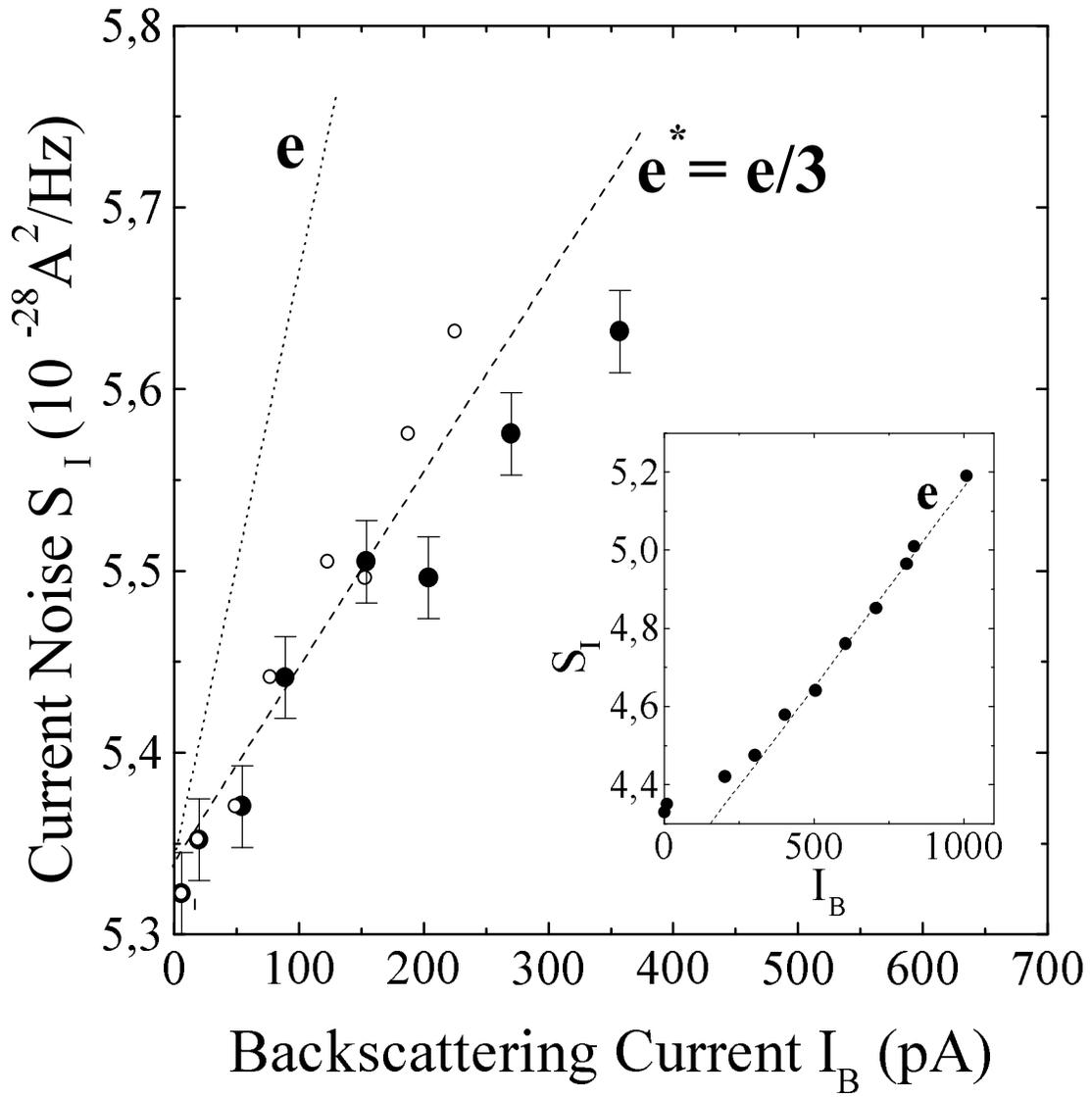
\begin{figure}[tbp]
\caption{Tunneling noise at $\nu=1/3$ ($\nu_{L}=2/3$) when
following path A and plotted versus $I_{B}=(e^2/3h)V_{ds}-I$ (filled circles)
and $I_{B}(1-R)$ (open circles). The slopes for $e/3$ quasiparticles (dashed
line) and electrons (dotted line) are shown. $\Theta=25mK$. Inset: data in same units
showing electron tunnelling for similar $G=0.32e^2/h$ but in the IQHE regime
($\nu_{L}=4$). The expected slope for electrons $2eI_{B}(1-R)$ ($R=0.68$, $%
I_{B}=(e^2/h)V_{ds}-I$) is shown. $\Theta=42mK$.}
\label{fig2}
\end{figure}

\begin{figure}[tbp]
\caption{Filled circles: Shot noise measured at $25mK$ versus $I_{B}$
corresponding to the paths B, C, and D of figure 1, and to a series of
measurements (E) at 150mK. Open circles: same data versus $I_{B}(1-R)$.}
\label{fig3}
\end{figure}

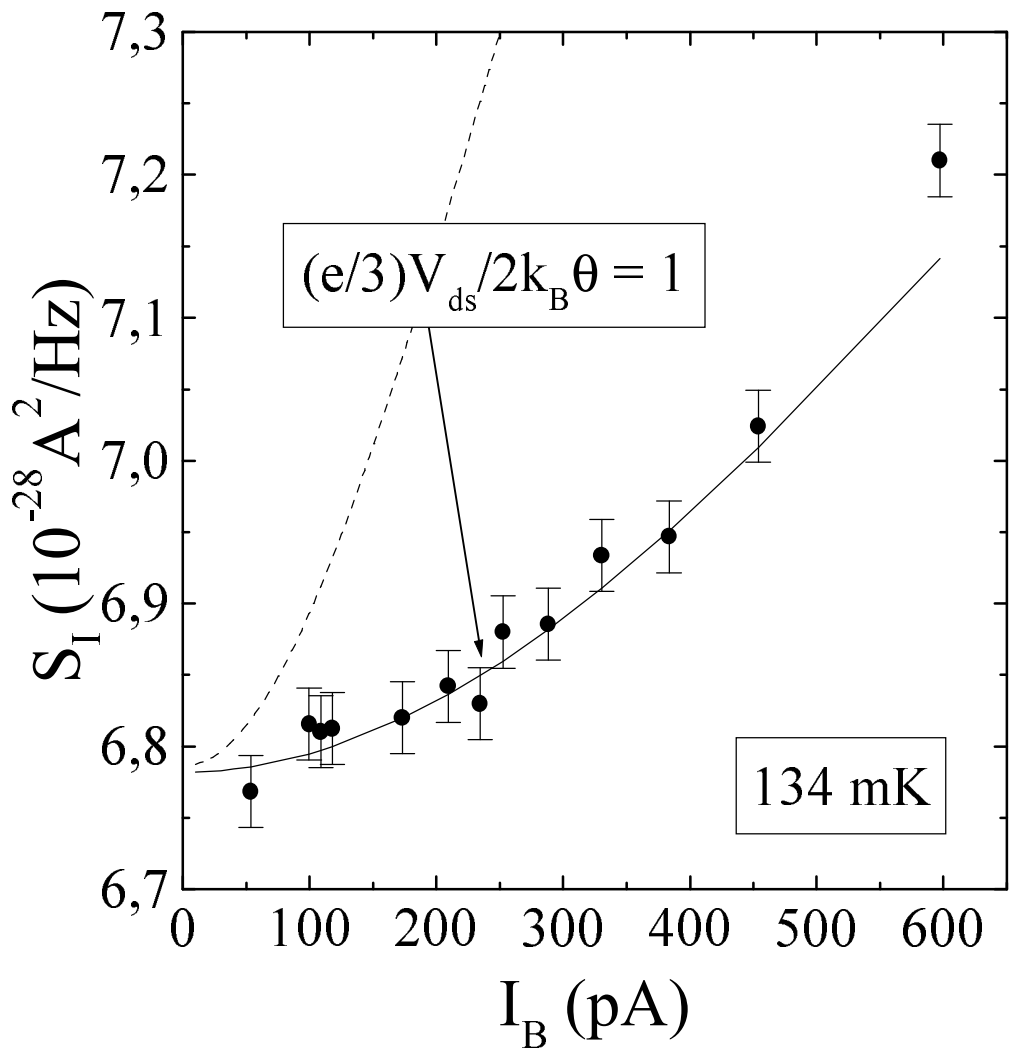
\begin{figure}[tbp]
\caption{ Cross-over from Johnson-Nyquist to shot noise. The arrow indicates
the data for which $e^{*}V_{ds}=2k_{B}\Theta$. A comparison with Eq.2 (solid
curve) and a similar expression for electrons (dotted curve) is shown.}
\label{fig4}
\end{figure}


\begin{references}
\bibitem{Note}  For incoherent tunneling $S_{I}\le 2e^{*}I_{B}$. This was
found when $\nu =\nu _{L}=1/3$. The weak QPC potential used to keep $%
I_{B}\ll I$ hardly compete with disorder and hopping occurs.
\end{references}
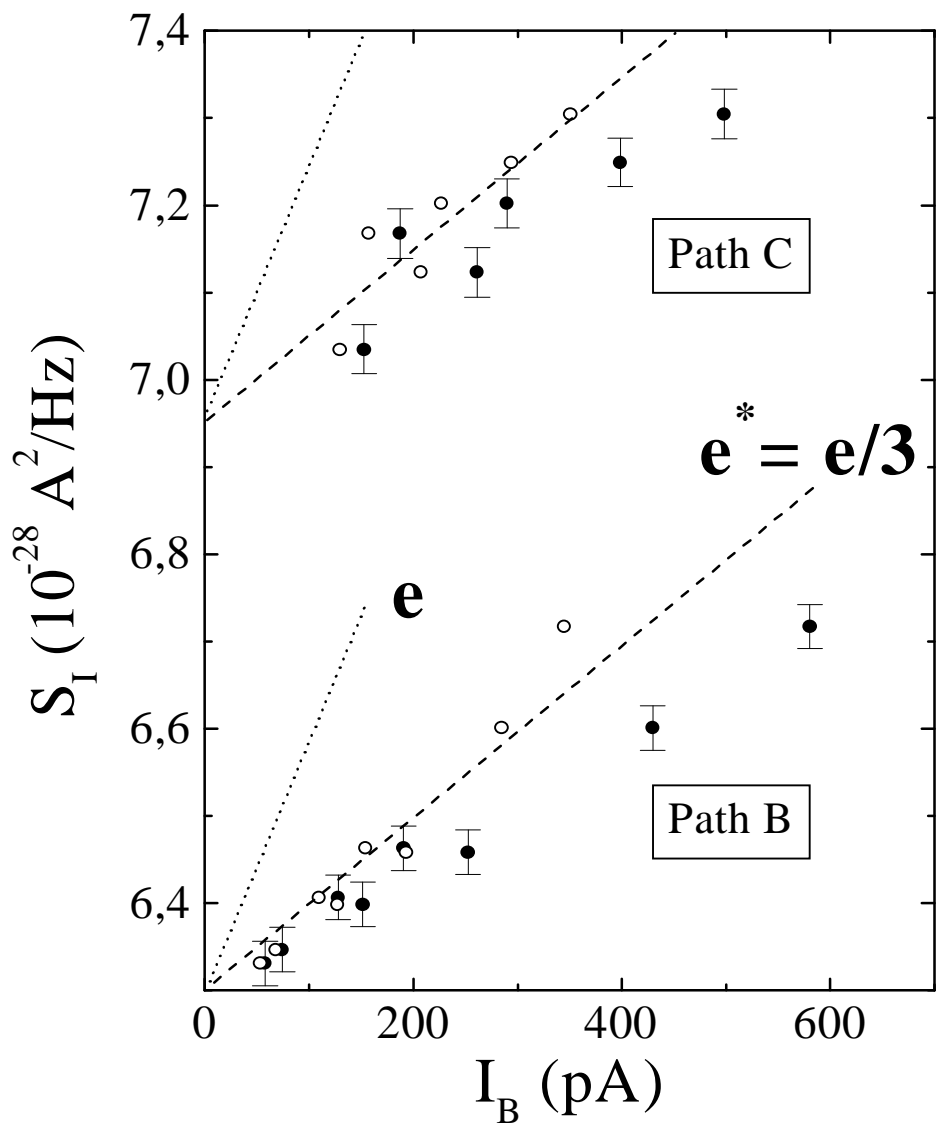
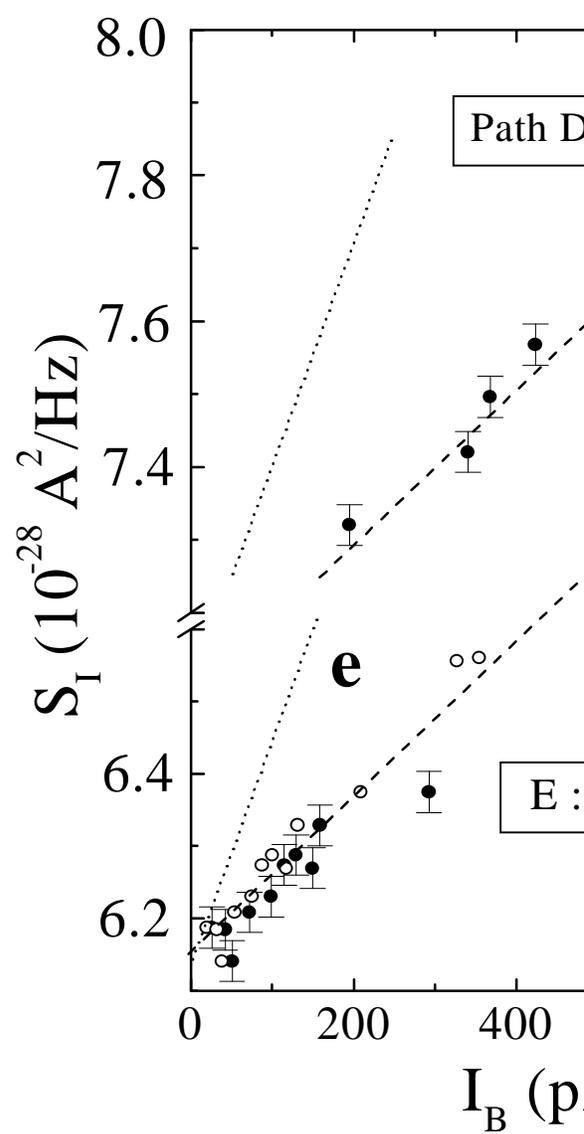
\end{document}